\documentclass{pasj00}

\begin{document}
\SetRunningHead{M.Serino et al.}{MAXI~J1421$-$613}

\title{Low-Mass X-Ray Binary MAXI~J1421$-$613 observed by MAXI GSC and
Swift XRT}

\author{
 Motoko~\textsc{Serino},\altaffilmark{1} %
%
 Megumi~\textsc{Shidatsu},\altaffilmark{2} %
 Yoshihiro~\textsc{Ueda},\altaffilmark{2} \\
 Masaru~\textsc{Matsuoka},\altaffilmark{1,3} %
 Hitoshi~\textsc{Negoro},\altaffilmark{4} %
 Kazutaka~\textsc{Yamaoka},\altaffilmark{5,6} %
 Jamie~\textsc{A.~Kennea},\altaffilmark{7} %
 Kosuke~\textsc{Fukushima},\altaffilmark{4} %
 Takahiro {\textsc Nagayama}\altaffilmark{8} %
}
\altaffiltext{1}
{MAXI team, Institute of Physical and Chemical Research (RIKEN), 2-1 Hirosawa, Wako, Saitama 351-0198, Japan}
\email{motoko@crab.riken.jp}
\altaffiltext{2}
{Department of Astronomy, Kyoto University, Kitashirakawa-Oiwake-cho, Sakyo-ku, Kyoto 606-8502, Japan}
\altaffiltext{3}
{ISS Science Project Office, Institute of Space and Astronautical Science (ISAS), Japan Aerospace Exploration Agency (JAXA), 2-1-1 Sengen, Tsukuba, Ibaraki 305-8505, Japan}
\altaffiltext{4}
{Department of Physics, Nihon University, 1-8-14 Kanda-Surugadai, Chiyoda-ku, Tokyo 101-8308, Japan}
\altaffiltext{5}
{Department of Particle Physics and Astronomy, Nagoya University, Furo-cho, Chikusa-ku, Nagoya, Aichi 464-8601, Japan}
\altaffiltext{6}
{Solar-Terrestrial Environment Laboratory, Nagoya University, Furo-cho, Chikusa-ku, Nagoya, Aichi 464-8601, Japan}
\altaffiltext{7}
{Department of Astronomy and Astrophysics, 0525 Davey Laboratory, Pennsylvania State University, University Park, PA 16802, USA}
\altaffiltext{8}
{Department of Physics and Astronomy, Kagoshima University, 1-21-35 Korimoto, Kagoshima 890-0065, Japan}


\maketitle

\begin{abstract}

Monitor of All sky X-ray Image (MAXI)
discovered a new outburst of an X-ray transient source named 
MAXI~J1421$-$613.
Because of the detection of three X-ray bursts from the source,
it was identified as a neutron star low-mass X-ray binary.
The results of data analyses of the MAXI GSC and the Swift XRT follow-up observations
suggest that
the spectral hardness
remained unchanged during the first two weeks of the outburst.
All the XRT spectra in the 0.5--10 keV band can be well explained by  
thermal Comptonization of multi-color disk blackbody emission. The
photon index of the Comptonized component is $\approx$ 2, which is 
typical of low-mass X-ray binaries in the low/hard state.
Since X-ray bursts have a maximum peak luminosity,
it is possible to 
estimate the (maximum) distance from its observed peak flux.
The peak flux of the second X-ray burst, which was observed by the GSC,
is about 5 photons cm$^{-2}$ s$^{-1}$.
By assuming a blackbody spectrum of 2.5 keV,
the maximum distance to the source is estimated as 7 kpc.
The position of this source is contained by the large error regions of
two bright X-ray sources detected with Orbiting Solar Observatory-7
(OSO-7) in 1970s.
Besides this, no past activities at the XRT position are reported in
the literature.
If MAXI~J1421$-$613 is the same source as (one of) them, the outburst
observed with MAXI may have occurred after the quiescence of 30--40
years.

\end{abstract}

\section{Introduction}

   Thermonuclear (type I) X-ray bursts are one of the key phenomena
   in understanding previously unknown X-ray sources.
   The cause of a sudden energy release is nuclear burning of
   accreted H/He fuel on the neutron star \citep{1993SSRv...62..223L}.
   Once an X-ray burst is detected, we know that the source
   must be a neutron star binary with a low-mass companion.
   Moreover, the existence of the characteristic luminosity of type I
   X-ray bursts 
   with photospheric radius-expansion (PRE) 
   is useful to determine the limit of the distance to
   these sources. 
   The luminosity of those
   bursts are thought to have reached the Eddington luminosity,
   while recent studies of PRE bursts have shown the variation of
   peak luminosity of $\sim$ 10\%
   \citep{1984PASJ...36..551E,2003A&A...399..663K,2008ApJS..179..360G}.
 
\begin{figure}
  \begin{center}
    \FigureFile(75mm,80mm){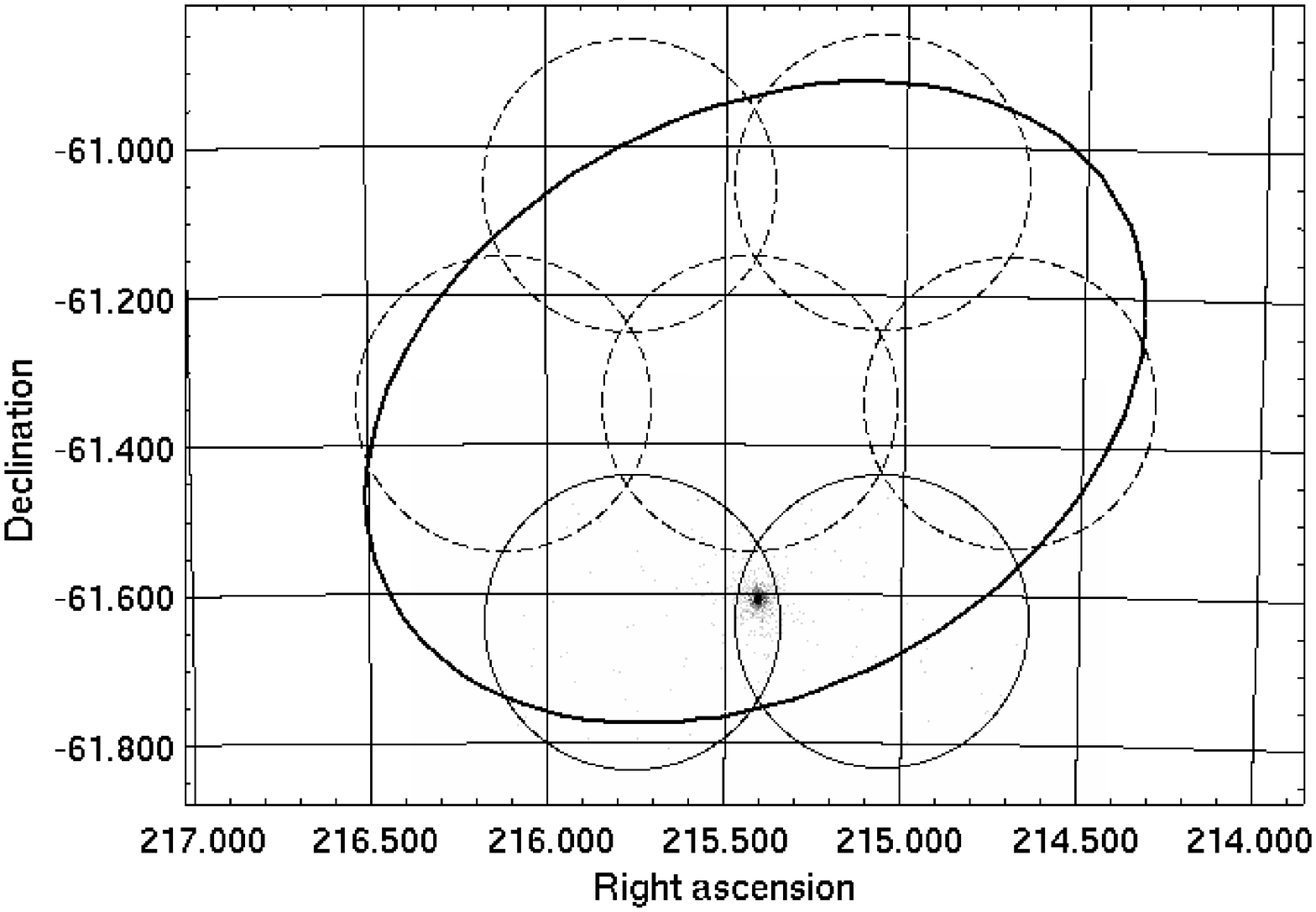}
  \end{center}
  \caption{An XRT image of MAXI~J1421$-$613 with the GSC error ellipse
  (large thick ellipse).
  Each solid and dashed circle corresponds to the XRT field of view
  of a pointing observation.
  The X-ray events accumulated by two pointing observations
  (shown with solid circles) are used for the image.}
  \label{fig:1}
\end{figure}

   \textit{The MAXI Nova-Alert System} 
   \citep{2010ASPC..434..127N} 
   triggered on a source at 01:13:07 UT on 2014 January 9.
   The position of the source was calculated as
   R.A. = 215.413 deg, Dec = $-$61.345 deg
   and reported to the Astronomer's Telegram
   as MAXI~J1421$-$613
   \citep{2014ATel.5750....1M}.
   At 19:35 on the same day, Swift performed a Target of Opportunity observation, tiling the MAXI GSC error ellipse with 7 pointings, 
   to confirm the detection. In the Swift X-ray Telescope 
   (XRT; \cite{2005SSRv..120..165B}) 
   data, there was an X-ray point source,
   in two out of the seven tiles
   \citep{2014ATel.5751....1K}.
   Figure \ref{fig:1} shows the image of MAXI~J1421$-$613 observed
   by Swift XRT 
   with the GSC error ellipse. 
   The position was determined by Swift XRT, utilizing UVOT to correct for astrometric errors \citep{PhilEvansPaper} as
   R.A., Dec(J2000) = 215.40838 deg, $-$61.60693 deg 
   \citep{2014ATel.5751....1K}.
   Utilizing 6ks of Photon Counting (PC) mode data 
   taken between January 14 and 20,
   \citet{2014ATel.5780....1K} revised the position to 
   R.A., Dec(J2000) = 215.40504 deg, $-$61.60700 deg. 
   The error radius for this revised position is 1.5 arcsec 
   (90\% confidence).
   It was also reported that the BAT Transient Monitor
   \citep{2013ApJS..209...14K} did not detect the source,
   suggesting that the source had a soft spectrum.
   Although the flux in the BAT energy band was low at that point,
   there may be an excess in the BAT light curve around January 14%
   \footnote{http://swift.gsfc.nasa.gov/results/transients/weak/MAXIJ1421-613/}.
   A radio counterpart of this source was detected by 
   the Australia Telescope Compact Array (ATCA)
   \citep{2014ATel.5759....1C,2014ATel.5802....1C}.
   The final position obtained by ATCA is 
   R.A., Dec (J2000) = \timeform{14h21m37s.2}, \timeform{-61D36'25''.4}
   with uncertainties of 0.3 arcsec in both RA and Dec.

   At least three type I X-ray bursts were detected
   from the source. 
   The first one was detected at 19:05 UT on January 10 
   by JEM-X on-board INTEGRAL \citep{2014ATel.5765....1B},
   the second one at 03:16 UT on January 16 by MAXI GSC, and
   the third one at 08:40 UT on January 18 by Swift
   BAT and XRT \citep{2014GCN..15749...1B}.
   The detection of the X-ray bursts reveals that the source is
   a low-mass X-ray binary containing a neutron star.

   In this paper, we analyze the data of MAXI~J1421$-$613 observed with
   MAXI GSC and Swift XRT. Section \ref{ss:2.1} presents the light
   curves and the results of spectral analyses covering the whole
   outburst period. We also report an upper flux limit obtained by a
   Suzaku observation. In section \ref{ss:xrb}, we focus on the X-ray
   bursts. The results of follow-up observations in near-infrared are
   presented in section \ref{ss:2.3}. In section \ref{ss:dist}, we discuss
   the distance to the source and constrain the spectral type of the
   companion star. The origin for an rapid decay observed at the end
   of the outburst is discussed in section \ref{ss:decay}. Finally, in
   section \ref{ss:catalog}, we discuss possible past activities of 
   MAXI~J1421$-$613, using X-ray source catalogs in the literature.

\section{Observations and Data Analyses}

\subsection{Light curve and spectra of the outburst}
 \label{ss:2.1}
   MAXI~J1421$-$613 has been monitored by MAXI 
   \citep{2009PASJ...61..999M}
   throughout the outburst.
   In addition, Swift XRT carried out pointed observations approximately every
   other day. We used these data sets to investigate the overall profile
   of the outburst.

\begin{figure}
  \begin{center}
    \FigureFile(80mm,80mm){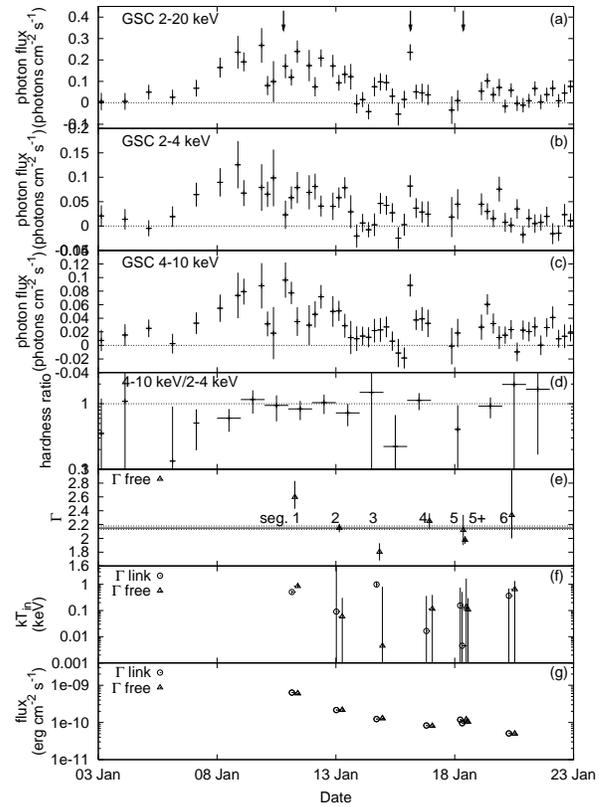}
  \end{center}
  \caption{The light curves (a, b, c) and the hardness ratio (d)
  observed by the GSC, and the spectral
  parameters of MAXI~J1421$-$613 derived by fitting
  the XRT data (e, f, g). The error bars on the GSC light curves
  and the hardness ratio are at 1$\sigma$. 
  The arrows in the top panel indicate the time of the
  X-ray bursts.
  The lower panels show the spectral parameters of the
  fit with {\tt nthComp} model. 
  The numbers in the panel (f) represent the seg.\ numbers.
  The asymptotic power-law indices $\Gamma$,
  the seed photon temperature, and the observed fluxes are
  shown in panel (e), (f), and (g), respectively.
  The bottom panel (g) shows the flux in the 0.5--10 keV band.
  If $\Gamma$ is linked over the segments, we obtained the parameters
  shown with circles. The best-fit value (and its error region) of linked $\Gamma$ 
  is shown with the solid (and dashed) horizontal lines.
  Otherwise, the obtained parameters are plotted with the triangles.
  The errors on these spectral parameters are
  at the 90 \% confidence level.
  }
  \label{fig:2}
\end{figure}

   The upper three panels (a, b, c) of figure \ref{fig:2} show the GSC
   light curves of MAXI~J1421$-$613 that are publicly available
   \footnote{http://maxi.riken.jp/top/index.php?cid=1\&jname=J1421-616}.
   The time bin size of the light curves are 6 h.
   The brightening of the source started around 
   January 7 and the flux increased almost linearly toward
   the peak on January 10. The peak photon flux in the 2--20 keV energy
   band was about 0.25 photons cm$^{-2}$ s$^{-1}$.
   Assuming the spectral model derived from the XRT observations
   ({\tt nthComp} of disk blackbody with $T_{\mathrm{in}} =$ 1.0 keV, 
   $\Gamma = 2.1$, $kT_{\mathrm{e}} = 20$ keV,
   with a photoelectric absorption of 
   $N\mathrm{_H}=4.8 \times 10^{22}$ cm$^{-2}$; see below),
   we estimated the flux 
   to be 2.6 $\times 10^{-9}$ ergs cm$^{-2}$ s$^{-1}$ in the same
   energy band.
   Then the flux decreased and went back to the
   background level on January 21. The three arrows in the top panel
   indicate the time of the X-ray bursts from the source.
   The GSC data point at the second arrow contains the emission of
   the X-ray burst. The fourth panel (d) shows the 
   hardness ratio between the 4--10 keV and 2--4 keV bands,
   which is binned up to 24 h.

   A series of monitoring observations (Target ID $=$ 33098)
   by Swift XRT started
   on January 11. There were eleven observation segments until 
   the last observation on February 4. Two of them
   (seg.\ 9 and 10) were performed with the PC mode and the rest were with
   the WT mode.
   An X-ray burst occurred during seg.\ 5, which triggered
   Swift BAT \citep{2014GCN..15749...1B}. 
   The automated follow-up observation with the PC mode started immediately
   after the burst. We also analyzed the data obtained by this
   follow-up observation (designate as seg.\ 5$+$ in this paper).
   At the time of seg.\ 7, the source had already become faint
   and the poor photon statistics did not allow us to perform
   spectral fitting. We thus used seven observations
   (seg.\ 1--6 and 5$+$) for the spectral analysis. 
   For seg.\ 5, we excluded the data of the last 40 s of the observation, 
   when the X-ray burst occurred.

   We extracted spectra from event data of each observation
   using XSELECT, and analyzed them with XSPEC. First, we tried 
   three simple models;
   power law with high energy exponential cutoff model ({\tt Cutoffpl}),
   power-law model ({\tt Powerlaw}),
   and blackbody model ({\tt Bbody}).
   All of them are multiplied by 
   the photo-electric absorption model 
   ({\tt Wabs}, \cite{1983ApJ...270..119M}).
   We jointly fit the data of the seven observations 
   with a common absorption column density.
   The chi-squared (and the degrees of freedom) of the fitting with 
   {\tt Cutoffpl}, {\tt Powerlaw}, and {\tt Bbody} models 
   are 1392.32 (1402), 1527.32 (1409), and 1591.70 (1409), 
   respectively. Therefore, {\tt Cutoffpl} model is the most favored model
   to describe the spectral shape of this source.
   If we link the photon indices of the {\tt Cutoffpl} model
   among all the segments, we still have an acceptable fit
   (the chi-squared is 1464.92 for 1408 degrees of freedom).
   In this case, the photon index is 0.6 $\pm$ 0.3 and the cutoff
   energy is $\sim 3$ keV for all the segments.
   Although the errors on the photon indices are large,
   the result is consistent with the supposition that 
   the hardness (or spectral shape) did not 
   change significantly throughout the outburst. 

   It is suggested that a hard-to-soft spectral transition occurs at 
   1--4\% of the Eddington luminosity
   \citep{2003A&A...409..697M,2012PASJ...64..128A}.
   If the X-ray burst observed by the GSC reached
   the Eddington luminosity (see section \ref{ss:dist}), the
   flux of 4\% of that becomes 
   $\sim 3 \times 10^{-9}$ ergs cm$^{-2}$ s$^{-1}$.
   The observed flux was lower than this value throughout the outburst,
   unlike other transient neutron star low-mass X-ray 
   binaries with spectral state transitions observed by MAXI 
   \citep{2012PASJ...64..128A,2013PASJ...65...58S}.
   Therefore, it is reasonable to interpret that MAXI~J1421$-$613 remained in 
   the hard state during the outburst.

   Then, we tried some typical spectral models for the hard state.
   A typical spectrum of the hard state consists of a soft thermal
   component, which is described as blackbody or multi-color disk blackbody
   ({\tt diskbb} \cite{1984PASJ...36..741M}), 
   and a hard Comptonized component
   \citep{2011MNRAS.416..637R,2012PASJ...64...72S,2012PASJ...64..112M}.
   First, we tried thermally Comptonized continuum model
   ({\tt nthComp}, \cite{1996MNRAS.283..193Z}) $+$ {\tt diskbb} with
   {\tt Wabs}. However, the parameters were not well constrained,
   perhaps due to the poor statistics or the low flux of the soft component. 
   Then, we simplified the
   model as {\tt nthComp} with {\tt Wabs}.
   Here we selected the disk blackbody model as the seed photons.
   Because the energy band is limited (0.5--10 keV), we were not able to
   determine the electron temperature $T_{\mathrm{e}}$ from the
   spectral fit. 
   Thus, we fixed $T_{\mathrm{e}}$ to 20 keV%
   \footnote{We also tried $T_{\mathrm{e}}$ of 40 keV but the result
   did not change significantly}. In addition, we fixed
   the column density of this fit to
   4.8 $\times 10^{22}$ cm$^{-2}$, which was derived from
   the spectral fit of the third X-ray burst (see section \ref{ss:xrb}),
   since the column density of X-ray burst spectra
   is not affected much by the uncertainty in the shape of
   the continuum.

\begin{figure}
  \begin{center}
    \FigureFile(80mm,80mm){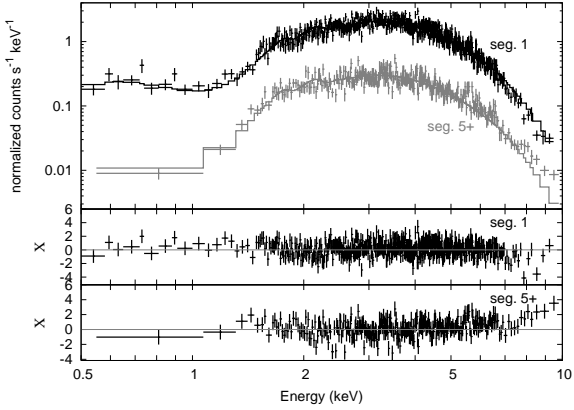}
  \end{center}
  \caption{The spectra of the XRT data of segs.\ 1 (black) and 
  5$+$ (gray).
  The data (crosses) and folded models (steps) are
  shown in the top panel and the residuals from the best fit 
  {\tt nthComp} with {\tt Wabs} models are shown in the middle and
  bottom panels.
  }
  \label{fig:spec}
\end{figure}

   Figure \ref{fig:spec} shows the samples of the spectra 
   of the seg.\ 1 (black) and 5$+$ (gray) observations fitted by
   {\tt nthComp} with {\tt Wabs} model.
   The spectral parameters are shown in figure \ref{fig:2}.
   The panel (e), (f), and (g) of figure \ref{fig:2} show 
   the asymptotic power-law indices $\Gamma$,
   the temperature ($kT_{\mathrm{in}}$) of the seed photons,
   and the flux in the 0.5--10 keV band, respectively.
   Although some parameters are not well constrained,
   the decay trend of the best fit fluxes agrees with the MAXI GSC 
   light curve.
   The index $\Gamma$ distributes around 2, which is consistent
   with the interpretation of the hard state.

   We requested a Suzaku observation of MAXI~J1421$-$613.
   The observation started at 12:20:40 UT on 2014 January 31,
   and the exposure was about 49 ks.
   The clock mode 
   of the X-ray Imaging Spectrometer (XIS)
   was normal and no window and burst options
   were applied. 
   The source was not significantly detected, however. We calculated
   an upper flux limit, using the XIS0 data of both 3x3 and 5x5 editing modes.
   We extracted the events from the central 2.5 arcmin region.
   The total observed count for the 48.4 ks exposure was 1030 counts
   in the 0.5--10 keV band.
   The background region of the same size was taken to the South of the
   source region, and the observed count for the same exposure
   and energy band was 1084 counts.
   Therefore, the 3$\sigma$ error on the observed count is 138 counts and
   the corresponding count rate is 0.0028 counts s$^{-1}$.
   Assuming the spectral parameters of the last XRT observation,
   ({\tt nthComp} of disk blackbody with $kT_{\mathrm{in}} =$ 0.37 keV, 
   $\Gamma = 2.1$, $kT_{\mathrm{e}} = 20$ keV,
   with a photoelectric absorption of 
   $N\mathrm{_H}=4.8 \times 10^{22}$ cm$^{-2}$), 
   we obtained the 3$\sigma$ upper limit of 
   1.2 $\times 10^{-13}$ ergs cm$^{-2}$ s$^{-1}$ (0.5--10 keV).
   This limit is consistent with the upper limit of the Chandra observation
   reported by \citet{2014ATel.5894....1C}.

\subsection{Light curves and spectra of the X-ray bursts}
  \label{ss:xrb}
\begin{figure}
  \begin{center}
    \FigureFile(80mm,80mm){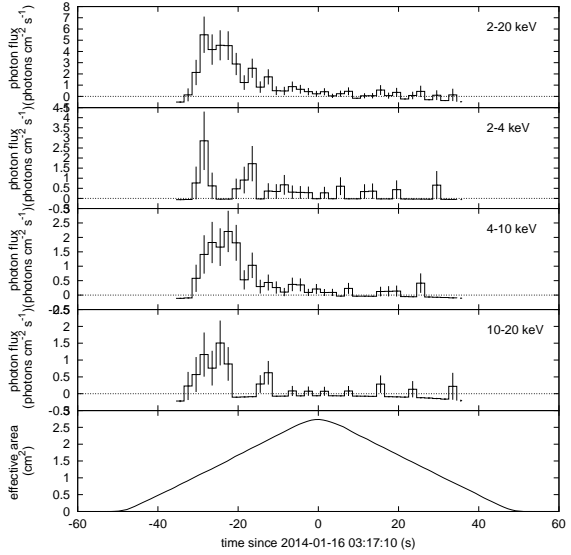}
  \end{center}
  \caption{The light curve of the second burst observed by the GSC.
  The photon fluxes were corrected for the effective area shown
  in the bottom panel.}\label{fig:3}
\end{figure}

   In order to understand the nature of the X-ray bursts, we analyzed the GSC and XRT data
   taken at the time of the bursts. Figure \ref{fig:3} shows the
   effective area corrected light curves of the GSC transit at 
   the X-ray burst.
   The burst occurred in the first half of the transit and lasted
   about 36 seconds.
   Although the statistics are not sufficient, there is a hint of
   spectral evolution. At the time around $-$25 s in figure \ref{fig:3},
   the photon flux in the 2--4 keV band decrease, while that of in the
   10--20 keV band increase, suggesting an increase of the temperature. 

\begin{figure}
  \begin{center}
    \FigureFile(80mm,80mm){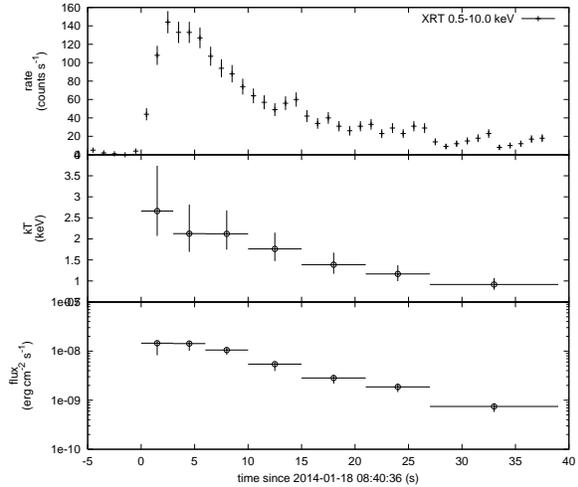}
  \end{center}
  \caption{The light curve and spectral parameters of 
  the third burst observed by the XRT.
  The top panel shows the observed count rate in the XRT 0.5--10.0 keV
  band. The middle and bottom panels show the observed temperature
  in keV and the observed flux in the 0.5--10.0 keV band derived from the
  joint spectral fitting.
  }\label{fig:4}
\end{figure}

   The third X-ray burst was detected during the pointed observation
   of the XRT. The top panel of figure \ref{fig:4} shows the XRT light curve
   in the 0.5--10 keV band. We performed time-resolved spectral
   analyses of the burst. Since the count rate is larger than 100
   counts s$^{-1}$, the first and second parts may be affected
   by pile-up%
   \footnote{XRT User's Guide\\
   (http://swift.gsfc.nasa.gov/analysis/xrt\_swguide\_v1\_2.pdf)}.
   Hence, we exclude photons in the central 2 pixels of the image
   for these intervals. 

   We fitted the spectra of seven time intervals, using the blackbody model
   with a photoelectric absorption.
   We performed a joint spectral analysis with a common absorption
   column density, which resulted in 
   ($4.8_{-1.1}^{+1.3}$) $\times 10^{22}$ cm$^{-2}$.
   The results of the spectral analyses are shown
   in the same figure. We find that the temperature decreases as the flux decreases, 
   which is typical of the decay part of X-ray bursts.

\subsection{Follow-up observations in near-infrared}
\label{ss:2.3}
Near-infrared photometric observations in the $J$ (1.25 $\mu$m), $H$
(1.63 $\mu$m), and $K_{\rm S}$ (2.14 $\mu$m) bands were carried out with
the SIRIUS camera \citep{ref_IRSF} on the 1.4 m telescope of Infrared
Survey Facility (IRSF) in South African Astronomical Observatory (SAAO)
on 2014 January 16, 17, 20, 21, 22, and 24.
The exposure in each night was 300 sec (after combining 15
dithered frames with a 20-sec exposure), and the typical seeing in the
$J$ band was $\approx$1''.5 in full width at half maximum. The standard
data reduction (bias correction, dark subtraction, flat fielding, sky
subtraction, and combining dithered images) was performed through the
IRSF pipeline software ``sirius09'' on the IRAF (Image Reduction and
Analysis Facilities; \cite{ref_IRAF}) package.

The counterpart of MAXI~J1421$-$613 in any band was not detected at any
night within the 90\% error circle (1''.5) of the Swift XRT position
\citep{2014ATel.5751....1K}. The upper limits of MAXI~J1421$-$613 
magnitudes were estimated from 3 standard deviations (3$\sigma$) of the 
averaged background counts in nearby source-free regions. Here, to obtain 
the tightest constraints, all the images are co-added by using ``pyIRSF'', 
a python package of the IRSF data reduction, after excluding those of 
January 20, which are out of focus, and those of January 22, where the
signal-to-noise ratios are twice worse than in the other data.
The conversion from the counts to
magnitudes was based on the comparison of stellar photometry around the
source position with the magnitudes of 2MASS all-sky catalog
\citep{ref_2MASS}. The apparent magnitudes were
$>$18.9 mag, $>$18.2 mag, and $>$16.9 mag in the $J$, $H$, $K_{\rm S}$
bands,
respectively. 

\section{Discussions}

\subsection{X-ray bursts and distance to the source}
\label{ss:dist}
   We calculate the bolometric peak flux of the two X-ray bursts
   observed by the GSC and the XRT. The peak photon flux
   of the second X-ray burst, observed by the GSC, is about
   5 photons cm$^{-2}$ s$^{-1}$ in the 2--20 keV band.
   By assuming the blackbody spectrum of 2.5 keV,
   the bolometric peak flux is calculated to be $7\times 10^{-8}$ 
   ergs cm$^{-2}$ s$^{-1}$.

   For the third X-ray burst observed by the XRT (figure \ref{fig:4}), 
   the peak flux (the first time interval) is 
   1.5 $\times 10^{-8}$ ergs cm$^{-2}$ s$^{-1}$ 
   in the 0.5--10 keV band.
   The (unabsorbed) bolometric flux based on the best fit model is
   3.2 $\times 10^{-8}$ ergs cm$^{-2}$ s$^{-1}$.

   Assuming the empirical maximum luminosity of X-ray bursts, 
   3.8 $\times 10^{38}$ ergs s$^{-1}$ 
   (for helium burning, \cite{2003A&A...399..663K}), 
   and using the peak burst flux of 
   7 $\times 10^{-8}$ ergs cm$^{-2}$ s$^{-1}$,
   we estimate the maximum distance to the source as 7 kpc.


By using this distance, together with the hydrogen column density, 
upper limits of absolute magnitudes
of the near-infrared counterpart can be calculated.
To derive the limits conservatively, we calculate the extinction in each band
from the 90\% upper limit of the averaged
hydrogen column density ($N_{\rm H} = 6.1 \times 10^{22}$ cm$^{-2}$)
obtained from the X-ray spectra during the third X-ray burst
(see Section \ref{ss:xrb}), using
the $N_{\rm H}$ versus reddening ($E(B-V)$) relation in
\citet{ref_NH_vs_EB-V} and the extinction curve in
\citet{ref_extinction} for $R_V (=A_V/E(B-V)) = 3.1$.

The resultant 3$\sigma$ upper limits of the extinction-corrected
absolute magnitudes are $-4.51$ mag ($J$ band), $-2.24$ mag ($H$ band),
and $-1.05$ mag ($K_{\rm S}$ band). These constraints suggest that the
companion of the MAXI~J1421$-$613 should be fainter than B3 star, if it
is a main sequence star (\cite{ref_mag_vs_type}). 
This is consistent with the identification of MAXI~J1421$-$613 with 
a low-mass X-ray binary, which can exhibit type-I X-ray bursts.

  \subsection{The decay of the outburst}
  \label{ss:decay}
   The flux limit derived from the Suzaku observation 
   ($<$1.2 $\times 10^{-13}$ ergs cm$^{-2}$ s$^{-1}$ in the 0.5--10 keV
   band at 3$\sigma$)
   is lower than
   an extrapolation of the flux observed by the XRT (\S\ref{ss:2.1}).
   Compared with the observed flux of 
   5.1 $\times 10^{-11}$ ergs cm$^{-2}$ s$^{-1}$ (0.5--10 keV)
   on January 20, the flux of the source had decayed by more than two orders
   of magnitude in $\sim$10 days. This decay rate is much faster than the
   observed decay curve of the flux measured with the XRT (figure \ref{fig:2}).
   Such rapid decay has been observed from other neutron star 
   low-mass X-ray binaries. For example, in the 1997 February-March
   outburst of Aquila X-1, the luminosity decreased from $\sim 10^{36}$ 
   to $\sim 10^{33}$ ergs s$^{-1}$ in 10 days
   \citep{1998ApJ...499L..65C}.
   This decay is interpreted as the onset of the propeller effect
   \citep{2013ApJ...773..117A}.
   In the case of MAXI~J1421$-$613, the propeller
   effect is a possible cause for the sudden decrease of the source flux.

\subsection{Cataloged X-ray Sources at the Position of MAXI~J1421$-$613}
  \label{ss:catalog}
   At the revised position determined by the Swift XRT, there are 
   no previously cataloged X-ray sources whose position errors are less than 1
   arcmin.
   We find, however, that two X-ray sources reported in the literature
   contain the XRT position of MAXI~J1421$-$613 in their large error
   regions.

   The first one is an (anonymous) X-ray source
   reported by \citet{1975IAUC.2761S...1W} (hereafter W75),
   which was detected with OSO-7 between 1971 and 1972.
   The position is
   R.A., Dec (equinox 1972.0) = 
     \timeform{14h12m}, \timeform{-62D},
   with an uncertainty of $\sim$3 deg \citep{1977ApJ...218..801M}.
   The other cataloged source is MX~1418$-$61
   (or MXB~1418$-$61 
\footnote{
Although the identifier is ``MXB', 
the original reference of this source 
is not ``X-ray bursters and the X-ray sources of the 
galactic bulge'' \citep{1981SSRv...28....3L} and no X-ray bursts
are reported from this source in the literature.
}
in SIMBAD; \cite{2000A&AS..143....9W}),
   which was also detected with OSO-7 between 1971 and 1974
   \citep{1975IAUC.2765Q...1M,1977ApJ...218..801M}.
   The position and its error of MX~1418$-$61 was reported as
   R.A., Dec (B1950) = 
     \timeform{14h18m.5} $\pm$ \timeform{1m.8}, 
                    \timeform{-61D.4} $\pm$ \timeform{0D.2}.
   The XRT position is apart from the center of the error box by
   $\sim$ 4.6 arcmin but is within the error box. This source 
   might be the same one as (or a part of) the object reported by W75.

   Since the late '70s, there were several X-ray satellites
   that were able to monitor bright X-ray sources by covering a wide
   area of the sky,
   such as Hakucho ('79--'85), Ginga ('87--'91), 
   and RXTE ('95--2012).
   Although they discovered many X-ray transients including
   X-ray bursters, none of the cataloged X-ray sources is consistent
   the position of MAXI~J1421$-$613. 
   If MAXI~J1421$-$613 exhibited X-ray bursts,
   they could have been well detected with these observatories.
   The ROSAT All Sky Survey detected no X-ray source
   at the XRT position \citep{1999A&A...349..389V,2000IAUC.7432R...1V}, either.
   These facts imply that the
   source had been in the quiescent state for more than 30 years.
   Hence, if MAXI~J1421$-$613 is 
   identical with MX~1418$-$61 (and/or the object
   reported in W75), 
   the outburst of MAXI~J1421$-$613 may have occurred after the
   quiescence of 30--40 years.

\section{Conclusion}
   MAXI GSC triggered on a new outburst of the X-ray source
   MAXI J1421$-$613 and Swift XRT located its position with an accuracy 
   of 1.5 arcsec (90\% confidence).
   There were three X-ray bursts 
   during the outburst, which showed the source to be a
   low-mass X-ray binary containing a neutron star.
   All seven spectra of the XRT in the 0.5--10 keV band can be well explained by  
   thermal Comptonization of multi-color disk blackbody emission. 
   These results may suggest
   that the spectral state of this source remained the hard state.
   This behavior is consistent with the relation between the luminosity
   and the spectral transition. The fast decay at the end of outburst
   can be explained as the onset of the propeller effect.

   The X-ray burst observed
   by the GSC was brighter than that observed by the XRT. Using the peak
   flux of the burst, we estimated an upper-limit of the distance to the
   source as 7 kpc.
   Assuming the distance and the
   hydrogen column density determined by the analysis of the X-ray
   burst, the upper limits for the absolute magnitudes are -4.51, -2.24,
   and -1.05 mag for J, H, and Ks band, respectively.  It means the
   companion should be fainter than B3 star, and it is consistent with
   the identification of MAXI~J1421$-$613 with a low-mass X-ray binary.

   The position of MAXI~J1421$-$613 is consistent with MX~1418$-$61
   and the object reported in W75, both detected by OSO-7 in 1970s, within
   their very large positional errors ($\sim$0.4--3 deg). 
   Besides this, no past activities at the XRT position are reported in
   the literature.
   If MAXI~J1421$-$613
   is the same source as (one of) them, the outburst observed with MAXI
   may have occurred after the quiescence of 30--40 years.


\bigskip


This research has made use of the MAXI data provided by RIKEN, JAXA and 
the MAXI team, and Swift XRT data obtained from The High Energy 
Astrophysics Science Archive Research Center of NASA.
We thank the Suzaku operation team for arranging and carrying out 
the TOO observations. 
We thank Robert J. Czanik for the IRSF observations. This publication
made use of data products from the Two Micron All Sky Survey, which is a
joint project of the University of Massachusetts and the Infrared
Processing and Analysis Center/California Institute of Technology,
funded by the National Aeronautics and Space Administration and the
National Science Foundation.
This research was partially supported by the Ministry of Education, 
Culture, Sports, Science and Technology (MEXT), Grant-in-Aid 
No. 24740186.

\bibliographystyle{aa}
\bibliography{maxij1421}

\end{document}